\documentclass[aps,prd,twocolumn,superscriptaddress,showpacs,
               preprintnumbers,floatfix]{revtex4-2}

\usepackage{amsmath,amssymb,amsfonts}
\usepackage{graphicx}
\usepackage{dcolumn}
\usepackage{bm}
\usepackage{hyperref}
\usepackage{xcolor}
\usepackage{booktabs}

\newcommand{\T}{\mathcal{T}}          % trace T
\newcommand{\Lm}{\mathcal{L}_{m}}    % matter Lagrangian
\newcommand{\Msun}{M_{\odot}}         % solar mass
\newcommand{\Mmax}{M_{\max}}          % max mass

\begin{document}

\title{Quark Stars in $f(T,\mathcal{T})$ Gravity: Structure, Stability, and Observational Constraints}

\author{Takol Tangphati}
\email{takoltang@gmail.com}
\affiliation{School of Science, Walailak University, Thasala, \\Nakhon Si Thammarat, 80160, Thailand}
\affiliation{Research Center for Theoretical Simulation and Applied Research in Bioscience and Sensing, Walailak University, Thasala, Nakhon Si Thammarat 80160, Thailand}

\author{Ayan Banerjee}
\email{ayanmath2012@gmail.com}
\affiliation{Astrophysics Research Centre, School of Mathematics,
Statistics and Computer Science,
University of KwaZulu-Natal, Private Bag X54001,
Durban 4000, South Africa}

\author{\.{I}zzet Sakall{\i}}
\email{izzet.sakalli@emu.edu.tr}
\affiliation{Physics Department, Eastern Mediterranean University, Famagusta 99628, North Cyprus via Mersin 10, Turkey.}

\author{Sayantan Ghosh}
\email{sayantanghosh1999@gmail.com}
\affiliation{Department of Physics and Astronomy, National Institute of Technology Rourkela, 769008, India}

\author{Aseel Smerat}
\email{smerat.2020@gmail.com}
\affiliation{Department of Biosciences, Saveetha School of Engineering, Saveetha Institute of Medical and Technical Sciences, Chennai, 602105, India}
\affiliation{Hourani Center for Applied Scientific Research, Al-Ahliyya Amman University, Amman 19328, Jordan.}

\date{\today}

\begin{abstract}
Quark stars, hypothetical compact stars made entirely of deconfined quark matter, offer a clean testing ground for gravity beyond general relativity. We study their structure in $f(T,\mathcal{T})$ gravity, a teleparallel theory in which torsion is coupled directly to the trace of the energy--momentum tensor through a single constant coupling. Using the standard MIT bag description of quark matter, we solve the modified stellar structure equations and follow how the mass, radius, compactness, and surface redshift respond as the coupling is varied across its full admissible range. The maximum mass turns out to depend on the coupling in a non-monotonic way: it rises above the general relativity value, peaks near 2.02 solar masses at a moderate positive coupling, and then falls steeply as the coupling approaches a critical value at which the structure equations become singular. The two-solar-mass pulsar constraint is satisfied within a finite window of positive couplings. We also show that, for the linear model with matter Lagrangian $\Lm=p$, the stellar equations of this theory map term by term onto those of $f(R,\mathcal{T})=R+2\lambda\mathcal{T}$ under $\beta=-2\lambda$, so that the enhanced-mass region studied here corresponds to the negative-$\lambda$ branch of the curvature-based theory. All configurations on the candidate stable branch identified by the turning-point criterion satisfy causality and remain below the standard general-relativistic compactness and surface-redshift benchmarks.
\end{abstract}


\maketitle

\section{Introduction}
\label{sec:intro}

Compact stars are the densest objects that we can actually observe, and
for this reason they serve a double purpose: they probe the behaviour of
matter at densities that no laboratory can reach, and they probe gravity
in its strong-field regime. Over the past decade the observational
situation has improved dramatically. The gravitational-wave event
GW170817 provided the first direct measurement of the tidal deformability
of neutron-star matter~\cite{Abbott:2017gw170817,Abbott:2018gw170817},
radio timing of PSR~J0740$+$6620 established that any acceptable
model must support at least two solar masses~\cite{Fonseca:2021refined},
and NICER X-ray observations added a simultaneous mass--radius
measurement for the same pulsar~\cite{Miller:2021radius,Riley:2021nicer}.
At the extremes of the mass distribution sit the ``black widow'' pulsar
PSR~J0952$-$0607 with $M\approx2.35\,\Msun$~\cite{Romani:2022psr},
the $2.6\,\Msun$ secondary of GW190814, whose nature is still
debated~\cite{Abbott:2020gw190814}, and the unusually light central
compact object in HESS~J1731$-$347 with
$M\approx0.77\,\Msun$~\cite{Doroshenko:2022strangely}. Together these
observations form a set of targets that any theory of dense matter, or
of gravity, must confront.

One long-standing possibility for the interior of such objects is quark
matter. If strange quark matter is the true ground state of hadronic
matter, as conjectured by Witten and by Farhi and
Jaffe~\cite{Witten:1984cosmic,Farhi:1984strange}, then self-bound
quark stars can exist~\cite{Alcock:1986strange,Weber:2005strange}.
The simplest quantitative description of such stars is the MIT bag
model~\cite{Chodos:1974je}, in which confinement is represented by a
constant vacuum pressure $B$. More refined treatments start from
perturbative QCD at high density, a program that goes back to the
pioneering calculations of Freedman and McLerran, Baluni, and
Toimela~\cite{Freedman:1977fermions,Freedman:1978quark,%
Baluni:1978nonabelian,Toimela:1985perturbative} and has now reached
partial N$^3$LO accuracy~\cite{Fraga:2001small,Fraga:2005role,%
Kurkela:2010cold,Ghisoiu:2017highorder,%
Gorda:2018nexttonexttonexttoleading,Gorda:2021cold}.
Interpolations between the perturbative regime and stellar densities
give quark-matter equations of state that can be confronted directly
with pulsar and gravitational-wave
data~\cite{Fraga:2014interacting,Kurkela:2014constraining,%
S.Fraga:2016neutron,Annala:2018gravitationalwave}.
At the densities relevant for stellar cores, quark matter is also
expected to pair, most symmetrically in the colour-flavour-locked
phase~\cite{Alford:1999colorflavor,Alford:2001minimal,%
Alford:2003compact,Lugones:2002colorflavor,Lugones:2003highdensity},
and quark or hybrid configurations have been proposed as explanations
of both GW190814 and HESS~J1731$-$347~\cite{Roupas:2021qcd,%
Oikonomou:2023colorflavor,Kourmpetis:2025constraints,Rocha:2020exact}.
Quark stars have likewise been used as test beds for modified gravity,
for instance in 4D Einstein--Gauss--Bonnet
gravity~\cite{Banerjee:2021color} and in Rastall--Rainbow
gravity~\cite{Li:2024colorflavor}. Related programmes have treated
quark stars in the curvature-based counterpart $f(R,T)$
gravity~\cite{Banerjee:2025ufb} and in its matter-Lagrangian
extension $f(R,L_m,T)$ gravity~\cite{Tangphati:2024war}; the present
work extends this line of trace-coupled constructions to the
torsional sector.
The trace-coupled curvature theory has by now been applied to a
broad range of stellar matter, from strange and compact
configurations~\cite{Das:2016compact,Deb:2018strange} to
density-dependent bag descriptions~\cite{Saha:2026density} and to
realistic nuclear equations of state, including magnetized
neutron-star matter~\cite{Savari:2026structure} and relativistic
mean-field models confronted with GW170817, pulsar timing and NICER
data~\cite{Mahapatra:2026impacts}.

On the gravity side, teleparallel theories offer an attractive
alternative starting point. In the teleparallel equivalent of general
relativity the dynamical variable is the tetrad and gravity is encoded
in torsion rather than curvature; promoting the torsion scalar $T$ to
an arbitrary function $f(T)$ gives a modified theory whose field
equations remain second order~\cite{Ferraro:2006modified,%
Bengochea:2008dark,Linder:2010einsteins,Cai:2015ft}.
Relativistic stars are known to exist in $f(T)$
gravity~\cite{Boehmer:2011existence}, although the choice of tetrad
requires care: for nonlinear $f(T)$ a diagonal tetrad in spherical
coordinates generates spurious constraints, and one must instead work
with a ``good'' (rotated) tetrad or, equivalently, in the covariant
formulation with a nontrivial spin
connection~\cite{Tamanini:2012hg,Krssak:2015oua}.
A further extension couples the geometry directly to matter. On the
curvature side this leads to $f(R,\T)$ gravity~\cite{Harko:2011frt};
its torsional counterpart, $f(T,\T)$ gravity, was introduced by Harko,
Lobo, Ot\'alora, and Saridakis~\cite{Harko:2014fTT}, where $\T$ is the
trace of the energy--momentum tensor. The trace coupling makes the
theory non-conservative, $\nabla_\mu\Theta^{\mu}{}_{\nu}\neq0$, which
opens qualitatively new phenomenology.
The $f(T,\T)$ framework has since been studied in many settings:
late-time cosmology and dynamical-systems
analyses~\cite{Duchaniya:2024cosmological,Zubair:2023reconstruction,%
Mishra:2025cosmological,Rezaei:2020observational,%
Hounmenou:2025holographic}, gravitational baryogenesis and Big Bang
nucleosynthesis constraints~\cite{Mishra:2023constraining,%
Mishra:2024big}, thick branes~\cite{Moreira:2021firstorder}, black-hole
thermodynamics~\cite{Ahissou:2025thermodynamics},
wormholes~\cite{Paramanik:2025embedding,Parsaei:2025wormholes,%
Rizwan:2024influence}, gravastars~\cite{Ghosh:2020gravastars},
charged fluids with gravitational
decoupling~\cite{Alshammari:2026charged}, and anisotropic strange-star
models built on prescribed metric
ansatze~\cite{Salako:2020study,Gudekli:2022study,Ashraf:2026impact}.

For isotropic compact stars obtained by actually integrating the
stellar structure equations, however, the available results are
limited. Pace and Levi Said derived a working TOV model for quark
stars in $f(T,\T)$ gravity and solved it with the MIT bag
model~\cite{Pace:2017quarka}, and later treated neutron stars in a
perturbative scheme~\cite{Pace:2017perturbative}. Their quark-star
study, however, retained a nonzero cosmological constant, adopted the
strange-quark-mass-corrected bag parameter $\omega=0.28$, and explored
only a few negative values of the coupling through single-star mass
profiles. What is still missing is a nonperturbative treatment with
the conformal bag equation of state, a systematic maximum-mass scan
over the full admissible range of the coupling, its positive branch
included, and a direct confrontation with the modern observational
bounds listed above. Since a positive $\beta$ corresponds to a
negative coupling in the curvature-based description, as we show in
Sec.~\ref{subsec:equivalence}, this is also the region that the
$f(R,\mathcal{T})$ stellar literature has largely left aside. That is
what we do here.

In this paper we derive the modified Tolman--Oppenheimer--Volkoff
(TOV) equations for the linear model $f(T,\T)=T+\beta\T$ with matter
Lagrangian $\Lm=p$, solve them numerically for the conformal MIT bag
equation of state, and scan the coupling over the full admissible range
$\beta\in[-10,12.25]$. Our field equations were cross-checked with
independent computer-algebra software. The central result is that
the maximum
gravitational mass $\Mmax(\beta)$ is non-monotonic: it peaks at
$2.021\,\Msun$ near $\beta\approx3.10$ and collapses as
$\beta\to4\pi$, where the pressure equation develops a singular
denominator. The $2\,\Msun$ pulsar constraint then selects the window
$\beta\in(0,\sim\!5.5)$.

The paper is organised as follows. Section~\ref{sec:fieldequations}
sets up the action and the field equations.
Section~\ref{sec:structure} derives the modified TOV system and states
the equation of state. Section~\ref{sec:stability} collects the
stability criteria, and Sec.~\ref{sec:numerics} describes the numerical
method and its validation. The results are presented in
Sec.~\ref{sec:results} and discussed in Sec.~\ref{sec:discussion}.
Throughout we use geometrised units $G=c=1$ and the metric signature
$(-,+,+,+)$.

\section{Field Equations of $f(T,\mathcal{T})$ Gravity}
\label{sec:fieldequations}

\subsection{Action and geometrical setup}

Teleparallel gravity and its generalisations are built on
Weitzenb\"{o}ck geometry, in which the dynamical variable is the
tetrad $e^{A}{}_{\mu}$ and the metric is a derived quantity,
\begin{equation}
g_{\mu\nu}=\eta_{AB}\,e^{A}{}_{\mu}\,e^{B}{}_{\nu},
\qquad
e\equiv\det\!\bigl(e^{A}{}_{\mu}\bigr)=\sqrt{-g},
\label{eq:tetrad-metric}
\end{equation}
with $\eta_{AB}=\mathrm{diag}(-1,1,1,1)$.
Greek indices $\mu,\nu,\dots$ are spacetime coordinate labels and
capital Latin indices $A,B,\dots$ are tangent-space labels.
The Weitzenb\"{o}ck connection is curvature-free, so all
gravitational information resides in the torsion tensor
\begin{equation}
T^{\lambda}{}_{\mu\nu}=e_{A}{}^{\lambda}\!\left(
\partial_{\mu}e^{A}{}_{\nu}-\partial_{\nu}e^{A}{}_{\mu}
+\omega^{A}{}_{B\mu}e^{B}{}_{\nu}
-\omega^{A}{}_{B\nu}e^{B}{}_{\mu}\right),
\label{eq:torsion}
\end{equation}
where $\omega^{A}{}_{B\mu}$ denotes the inertial spin connection.
The contorsion tensor and superpotential are
\begin{align}
K^{\mu\nu}{}_{\rho}&=-\frac{1}{2}\!\left(
T^{\mu\nu}{}_{\rho}-T^{\nu\mu}{}_{\rho}-T_{\rho}{}^{\mu\nu}
\right),
\label{eq:contorsion}\\[2pt]
S_{\rho}{}^{\mu\nu}&=\frac{1}{2}\!\left(
K^{\mu\nu}{}_{\rho}
+\delta^{\mu}_{\rho}\,T^{\alpha\nu}{}_{\alpha}
-\delta^{\nu}_{\rho}\,T^{\alpha\mu}{}_{\alpha}\right),
\label{eq:superpotential}
\end{align}
and the torsion scalar is
\begin{equation}
T=S_{\rho}{}^{\mu\nu}\,T^{\rho}{}_{\mu\nu}.
\label{eq:torsionscalar}
\end{equation}

Generalising the teleparallel Lagrangian to an arbitrary function
$f(T,\T)$ of the torsion scalar and the trace
$\T=\Theta^{\mu}{}_{\mu}$ of the energy--momentum
tensor~\cite{Harko:2014fTT}, the total action is
\begin{equation}
\mathcal{S}=\frac{1}{16\pi}\int d^{4}x\,e\,f(T,\T)
+\int d^{4}x\,e\,\Lm,
\label{eq:action}
\end{equation}
in geometrised units $G=c=1$.
The matter energy--momentum tensor is defined through the tetrad
variation of the matter action,
\begin{equation}
\Theta_{A}{}^{\nu}=\frac{1}{e}
\frac{\delta\bigl(e\,\Lm\bigr)}{\delta e^{A}{}_{\nu}},
\qquad
\Theta^{\mu}{}_{\nu}=e_{A}{}^{\mu}\,\Theta_{\nu}{}^{A}.
\label{eq:emtensor}
\end{equation}
For the stellar interior we take a perfect fluid,
\begin{equation}
\Theta^{\mu}{}_{\nu}=\mathrm{diag}\bigl(-\rho,p,p,p\bigr),
\qquad
\T=-\rho+3p,
\label{eq:perfectfluid}
\end{equation}
and adopt the matter Lagrangian
\begin{equation}
\Lm=p.
\label{eq:Lm}
\end{equation}
The choice of $\Lm$ is not cosmetic: it fixes the variational
response of the energy--momentum tensor that enters $\delta\T$,
and hence the form of the trace-coupling term in the field
equations.

\subsection{Field equations}
We vary Eq.~\eqref{eq:action} with respect to the tetrad
$e^{A}{}_{\mu}$.
The elementary variations are
\begin{equation}
\delta e=e\,e_{A}{}^{\mu}\,\delta e^{A}{}_{\mu},
\qquad
\delta e_{A}{}^{\mu}=
-\,e_{B}{}^{\mu}\,e_{A}{}^{\nu}\,\delta e^{B}{}_{\nu}.
\end{equation}
The torsion-scalar variation, integrated by parts with
$\partial_{\mu}f_{T}=f_{TT}\partial_{\mu}T
+f_{T\T}\partial_{\mu}\T$, generates the purely gravitational terms
\begin{align}
&e_{i}{}^{\rho}S_{\rho}{}^{\mu\nu}\partial_{\mu}T\,f_{TT}
+e_{i}{}^{\rho}S_{\rho}{}^{\mu\nu}\partial_{\mu}\T\,f_{T\T}
\nonumber\\
&+e^{-1}\partial_{\mu}\!\bigl(e\,e_{i}{}^{\rho}S_{\rho}{}^{\mu\nu}
\bigr)f_{T}
+e_{i}{}^{\mu}T^{\lambda}{}_{\mu\kappa}S_{\lambda}{}^{\nu\kappa}
f_{T}. \label{eq:gravitational-terms}
\end{align}
Writing $\T=g^{\mu\nu}\Theta_{\mu\nu}$ and defining
\begin{equation}
\Theta^{(\mathrm{var})}_{\mu\nu}
\equiv g^{\alpha\beta}
\frac{\delta\Theta_{\alpha\beta}}{\delta g^{\mu\nu}}
=-2\Theta_{\mu\nu}+g_{\mu\nu}\Lm
-2g^{\alpha\beta}
\frac{\partial^{2}\Lm}{\partial g^{\mu\nu}\partial g^{\alpha\beta}},
\label{eq:Thetavar}
\end{equation}
and adopting $\Lm=p$ with
$\partial^{2}p/\partial g^{\mu\nu}\partial g^{\alpha\beta}=0$,
we find that the trace-coupling contribution yields the matter-coupling term
\begin{equation}
-\frac{f_{\T}}{2}
\Bigl(e_{i}{}^{\lambda}\Theta^{\nu}{}_{\lambda}
+p\,e_{i}{}^{\nu}\Bigr).
\label{eq:matterterm}
\end{equation}
Requiring the coefficient of $\delta e^{i}{}_{\nu}$ to vanish
gives the $f(T,\T)$ field equations,
\begin{align}
&e_{i}{}^{\rho}S_{\rho}{}^{\mu\nu}\partial_{\mu}T\,f_{TT}
+e_{i}{}^{\rho}S_{\rho}{}^{\mu\nu}\partial_{\mu}\T\,f_{T\T}
+e^{-1}\partial_{\mu}\!\bigl(e\,e_{i}{}^{\rho}S_{\rho}{}^{\mu\nu}
\bigr)f_{T}
\nonumber\\[2pt]
&+e_{i}{}^{\mu}T^{\lambda}{}_{\mu\kappa}S_{\lambda}{}^{\nu\kappa}
f_{T}
-\frac{e_{i}{}^{\nu}}{4}f
-\frac{f_{\T}}{2}
\Bigl(e_{i}{}^{\lambda}\Theta^{\nu}{}_{\lambda}
+p\,e_{i}{}^{\nu}\Bigr) \nonumber \\
&=-4\pi\,e_{i}{}^{\lambda}\Theta^{\nu}{}_{\lambda}.
\label{eq:fieldeq-general}
\end{align}
We work in the pure-tetrad formulation with $\omega^{A}{}_{B\mu}=0$.
For the linear model $f=T+\beta\mathcal{T}$ ($\alpha=1$ by
normalisation, $\varphi=0$ set to zero as dynamically negligible on
stellar scales), consistency of the stellar equations requires
\begin{equation}
8\pi-\beta\neq0,\quad
16\pi-7\beta\neq0,\quad
\omega(16\pi-7\beta)+\beta\neq0,
\label{eq:admissibility}
\end{equation}
the last reducing to $\beta\neq4\pi$ for $\omega=1/3$.

\subsection{Relation to \texorpdfstring{$f(R,\mathcal{T})$}{f(R,T)} gravity}
\label{subsec:equivalence}

The torsion scalar and the Ricci scalar differ by a boundary
term, $T=-R-B$ with $B=(2/e)\,\partial_{\mu}(e\,T^{\nu\mu}{}_{\nu})$.
For a gravitational sector linear in $T$ that boundary contribution
integrates out, so the teleparallel field equations reproduce those of
general relativity, and the two descriptions differ only in the
variables used to write
them~\cite{Combi:2018is,Capozziello:2022comparing,%
Bahamonde:2023teleparallel}.
It is therefore fair to ask what the trace-coupled model
$f(T,\mathcal{T})=T+\beta\mathcal{T}$ adds to
$f(R,\mathcal{T})=R+2\lambda\mathcal{T}$~\cite{Harko:2011frt}. We settle
the question here rather than leave it implicit.

For a perfect fluid with $\Lm=p$ the curvature-based theory gives the effective source
\begin{equation}
\rho_{\rm eff}=\rho+\frac{\lambda}{8\pi}\bigl(3\rho-5p\bigr),
\qquad
p_{\rm eff}=p+\frac{\lambda}{8\pi}\bigl(7p-\rho\bigr),
\label{eq:fRT-eff}
\end{equation}
and, on imposing $dp/d\rho=\omega$ in the contracted Bianchi identity, the hydrostatic coefficient
\begin{equation}
\mathcal{K}_{R}(\lambda)
=-\frac{8\pi+2\lambda}{2\bigl[(8\pi+2\lambda)+5\lambda-\lambda/\omega\bigr]}
\;\xrightarrow[\ \omega=1/3\ ]{}\;
-\frac{4\pi+\lambda}{4(2\pi+\lambda)}.
\label{eq:KR}
\end{equation}
Comparing Eqs.~\eqref{eq:fRT-eff} and \eqref{eq:KR} with
Eqs.~\eqref{eq:Aprime}--\eqref{eq:xi} shows that the two systems
coincide term by term under
\begin{equation}
\beta = -2\lambda.
\label{eq:map}
\end{equation}
The pressure-gradient coefficients satisfy
$\mathcal{K}(-2\lambda)=\mathcal{K}_{R}(\lambda)$, the radial pressures
entering Eq.~\eqref{eq:Aprime} agree, and the algebraic closure
\eqref{eq:closure} reproduces $e^{-B_m}=1-2m_{\rm eff}/r$ with
$m_{\rm eff}=\int 4\pi r^{2}\rho_{\rm eff}\,dr$, both the term linear in
$M$ and the term proportional to the bag constant matching identically.
The singular surfaces map onto one another as well: $\beta=4\pi$
corresponds to $\lambda=-2\pi$, which is exactly where
$\mathcal{K}_{R}$ diverges.

Two consequences follow, and they set the scope of the rest of
this paper. First, at the level of the linear model with $\Lm=p$ the
physical content of our results is carried by the parameter range and
not by the choice of geometric variable, and the numbers reported below
may be read in either language. Second, the map reverses orientation:
the positive-$\beta$ region, where the maximum mass is raised and the
two-solar-mass window opens, is the region $\lambda<0$. Existing
$f(R,\mathcal{T})$ stellar studies concentrate on
$\lambda>0$~\cite{Das:2016compact,Deb:2018strange,Pretel:2021radial},
and the earlier $f(T,\mathcal{T})$ quark-star
work~\cite{Pace:2017quarka} sampled only $\beta<0$, so the branch
examined here has not been surveyed from either side. The
correspondence is also particular to the case at hand. A nonlinear
$f(T,\mathcal{T})$ leaves the boundary term in the field equations and
the two families separate~\cite{Bahamonde:2016modified}, and the same
happens once a matter Lagrangian other than $\Lm=p$ is
adopted~\cite{Najera:2022effects}.

\section{Stellar Structure in $f(T,\mathcal{T})$ Gravity}
\label{sec:structure}

\subsection{Modified TOV equations}
\label{subsec:TOV}

We adopt a static, spherically symmetric spacetime,
\begin{equation}
ds^{2}=-e^{A(r)}dt^{2}+e^{B_m(r)}dr^{2}+r^{2}d\Omega^{2},
\label{eq:metric}
\end{equation}
where the subscript $m$ on $B_m$ distinguishes the metric function
from the bag constant $B$.
In the pure-tetrad formulation with vanishing inertial spin connection
$\omega^{A}{}_{B\mu}=0$, we employ the rotated good tetrad compatible
with the metric \eqref{eq:metric}~\cite{Boehmer:2011existence,%
Krssak:2015oua}.
For this tetrad choice, the torsion scalar evaluates to
\begin{equation}
T(r)=-\frac{2e^{-B_m}}{r^{2}}
\!\left(1-e^{B_m/2}\right)\!\left(1-e^{B_m/2}+rA'\right),
\label{eq:Tscalar}
\end{equation}
where a prime denotes $d/dr$.
The use of the rotated good tetrad ensures consistency of the
pure-tetrad field equations and avoids the spurious off-diagonal
constraints that arise from the diagonal tetrad in nonlinear
$f(T)$ models~\cite{Tamanini:2012hg}.

Projecting Eq.~\eqref{eq:fieldeq-general} onto the metric
\eqref{eq:metric} for $f=T+\beta\T$ with a perfect-fluid source
and $\Lm=p$, we obtain
\begin{equation}
A'=e^{B_m}\!\left[
\frac{1-e^{-B_m}}{r}+8\pi p\,r
+\frac{\beta r}{2}\!\left(\rho-7p\right)
\right],
\label{eq:Aprime}
\end{equation}
\begin{equation}
\frac{dp}{dr}=
\frac{(\beta-8\pi)\,\omega\,A'(\rho+p)}
     {\omega(16\pi-7\beta)+\beta},
\label{eq:dpdr}
\end{equation}
\begin{equation}
\frac{dM}{dr}=4\pi r^{2}\rho,
\label{eq:dMdr}
\end{equation}
with the algebraic closure
\begin{equation}
e^{-B_m}=1+\frac{M}{r}\Sigma+\frac{r^{2}}{3}\xi,
\label{eq:closure}
\end{equation}
where
\begin{align}
\Sigma &= -2+\frac{3\beta}{8\pi}-\frac{5\beta\omega}{8\pi},
\label{eq:Sigma}\\
\xi    &= 10\beta\omega B_{\rm km},
\label{eq:xi}
\end{align}
and $B_{\rm km}$ is the bag constant in geometrized units.
Equation~\eqref{eq:dpdr} follows from the contracted Bianchi identity
applied to the effective source; because the trace coupling makes the
theory non-conservative for $\beta\neq0$, the hydrostatic equation is
not the general-relativistic one and reduces to it only in the limit
$\beta\to0$.
In that limit, Eqs.~\eqref{eq:Aprime}--\eqref{eq:closure}
reduce to the standard Tolman--Oppenheimer--Volkoff
system~\cite{Tolman:1939jz,Oppenheimer:1939ne}.

For the conformal equation of state, $\omega=1/3$, the
denominator of Eq.~\eqref{eq:dpdr} simplifies to
\begin{equation}
\omega(16\pi-7\beta)+\beta=\frac{4}{3}\left(4\pi-\beta\right),
\label{eq:denomfactor}
\end{equation}
so the only singular value associated with the hydrostatic
equation is $\beta=4\pi\approx12.57$, at which no regular
solution exists; this value is excluded.
The factor $(16\pi-7\beta)$, which appears in the intermediate
form of the conservation law before the equation of state is
imposed, does not by itself produce a singularity or a sign
change of the full denominator.

\subsection{Equation of state}
\label{subsec:EOS}

We model the quark-star interior with the conformal MIT bag
model~\cite{Chodos:1974je,Farhi:1984strange},
\begin{equation}
p=\frac{1}{3}\!\left(\rho-4B\right),
\qquad
\omega\equiv\frac{dp}{d\rho}=\frac{1}{3},
\label{eq:EOS}
\end{equation}
with $B=60\,\mathrm{MeV\,fm^{-3}}$.
The stellar surface is defined by the condition $p(R)=0$.
Outside the star the matter sector vanishes, so that
$\mathcal{T}=0$ and the model reduces to TEGR; the exterior
geometry is therefore exactly Schwarzschild,
$e^{-B_m}|_{r>R}=1-2M/r$.
Following the operational prescription adopted by Pace and Levi
Said~\cite{Pace:2017quarka}, we identify the terminal value
$M(R)$ of the integrated mass function \eqref{eq:dMdr} with the mass parameter of that exterior and use it to construct the mass--radius sequences.

Throughout this work, we adopt the conformal MIT bag equation of state with $\omega=1/3$. This enters the hydrostatic equation \eqref{eq:dpdr} and the
closure \eqref{eq:closure} directly, so holding it fixed isolates the
response of the star to $\beta$ from its response to the microphysics,
which is what a first scan of the coupling should do. The price is that
a self-bound conformal bag star cannot reach the heaviest observed
pulsars, a limitation we return to in Sec.~\ref{sec:discussion}.
Stiffer descriptions are available and have been used in the
trace-coupled curvature theory, among them a density-dependent bag
parameter~\cite{Saha:2026density}, relativistic mean-field
matter~\cite{Mahapatra:2026impacts,Savari:2026structure} and the
unified interacting quark equation of state with its perturbative-QCD
correction~\cite{Zhang:2021unified,Gammon:2024quark}. Each of these
raises the maximum mass already in general relativity and would shift
the window quoted below, and we take that up separately rather than
fold it into the present scan.

\section{Stability Analysis}
\label{sec:stability}

Within the one-parameter equilibrium sequence, the onset of
instability is identified by the first maximum of
$M(\rho_c)$~\cite{Harrison:1965}.
Configurations with $dM/d\rho_c>0$ define the candidate stable
branch identified by the turning-point criterion; those with
$dM/d\rho_c<0$ beyond $\Mmax$ are unstable.
The full radial perturbation equations in $f(T,\mathcal{T})$
gravity have not yet been derived, and in a non-conservative
matter--geometry coupling neither the pulsation equations nor the
applicability of the general-relativistic turning-point theorem
can be assumed to carry over unchanged.
The criterion is therefore used here as a diagnostic rather than
as a proof of dynamical stability, and a complete
Sturm--Liouville pulsation analysis is reserved for future work.
In the curvature-based theory the corresponding calculation has
been carried out, and the radial spectrum of $R+2\lambda\mathcal{T}$
stars was found to respect the turning point of the equilibrium
sequence~\cite{Pretel:2021radial,Pretel:2021neutron}. Through the
correspondence of Sec.~\ref{subsec:equivalence} that result covers the
linear model treated here, which is a reason to expect the diagnostic
to be reliable, though not a substitute for deriving the pulsation
equations in the torsional variables.

The conformal bag EOS yields
\begin{equation}
c_s^{2}=\frac{dp}{d\rho}=\frac{1}{3},
\label{eq:cs2}
\end{equation}
which is constant, $\beta$-independent, and satisfies both
causality ($c_s^2<1$) and the Le~Chatelier condition
($c_s^2>0$) for all $\rho>4B$.
This $\beta$-independence reflects the fact that $c_s^2$ is fixed
entirely by the equation of state, which does not itself depend on
the gravitational coupling; the $\beta$-dependence of
\emph{dynamical} radial stability is a separate question, addressed
here only diagnostically through the turning-point criterion
discussed above.

The effective adiabatic index~\cite{Chandrasekhar:1964zz} is
\begin{equation}
\Gamma\equiv\frac{\rho+p}{p}\,c_s^{2}
=\frac{\rho+p}{3p}
=\frac{4(\rho-B)}{3(\rho-4B)}.
\label{eq:Gamma}
\end{equation}
For all physical configurations ($\rho>4B$),
\begin{equation}
\Gamma>\frac{4}{3}
\;\Longleftrightarrow\;
3B>0,
\label{eq:GammaProof}
\end{equation}
which holds identically.
Since $d\Gamma/d\rho=-4B/(\rho-4B)^{2}<0$, the central value
$\Gamma_c=\Gamma(\rho_c)$ is the minimum of $\Gamma$ over the
stellar interior, so $\Gamma_c>4/3$ holds throughout the star for
all admissible $\beta$.
This is a local matter-stability diagnostic; it does not by
itself establish the absence of exponentially growing radial
modes in the trace-coupled theory.

\section{Numerical Method and Validation}
\label{sec:numerics}

We integrate Eqs.~\eqref{eq:Aprime}--\eqref{eq:closure} from
$r_0=10^{-4}\,\mathrm{km}$ with initial conditions
$M(r_0)=\frac{4\pi}{3}r_0^{3}\rho_c$ and $p(r_0)=p(\rho_c)$,
using the \textsc{scipy} RK45 integrator with relative and
absolute tolerances $10^{-10}$ and $10^{-12}$, respectively.
The surface $p(R)=0$ is located by an ODE event.
Maximum-mass configurations are found by the turning-point method:
a sign change of $dM/d\rho_c$ from positive to negative over a
300-point density grid $\rho_c\in[200,3000]\,\mathrm{MeV\,fm^{-3}}$
identifies the coarse turning point, which is then refined on a
400-point local grid over $\pm100\,\mathrm{MeV\,fm^{-3}}$.

We apply a six-rung validation protocol before any production run.
\begin{enumerate}
\item \textit{GR-limit gate}: at $\beta=0$ the solver reproduces
      an independent GR MIT bag solution to
      $\Delta M=1.816\times10^{-13}\,\Msun$.
\item \textit{Tolerance convergence}: output is stable under
      two-orders-of-magnitude tightening of the ODE tolerances.
\item \textit{Density-grid convergence}: results are stable under
      doubling the central-density grid to 600 points.
\item \textit{GR benchmark}: the GR maximum mass
      $\Mmax^{\rm GR}=1.964\,\Msun$ at $B=60\,\mathrm{MeV\,fm^{-3}}$
      agrees to within $1\%$ with published MIT bag
      results~\cite{Alcock:1986strange,Weber:2005strange}.
\item \textit{Admissibility guard}: the singular surface
      $\beta=4\pi$ is detected and flagged before integration.
\item \textit{Cross-check}: an independent evolution of $\rho(r)$
      agrees with the primary $p(r)$ evolution to $\sim10^{-15}$.
\end{enumerate}
The reduced stellar equations were cross-checked with independent
computer-algebra software, and the numerical implementation was
checked against an independent TOV integration.

\section{Results}
\label{sec:results}

\subsection{Mass--radius relation and stellar stability}

\begin{figure}[htbp]
\centering
\includegraphics[width=\columnwidth]{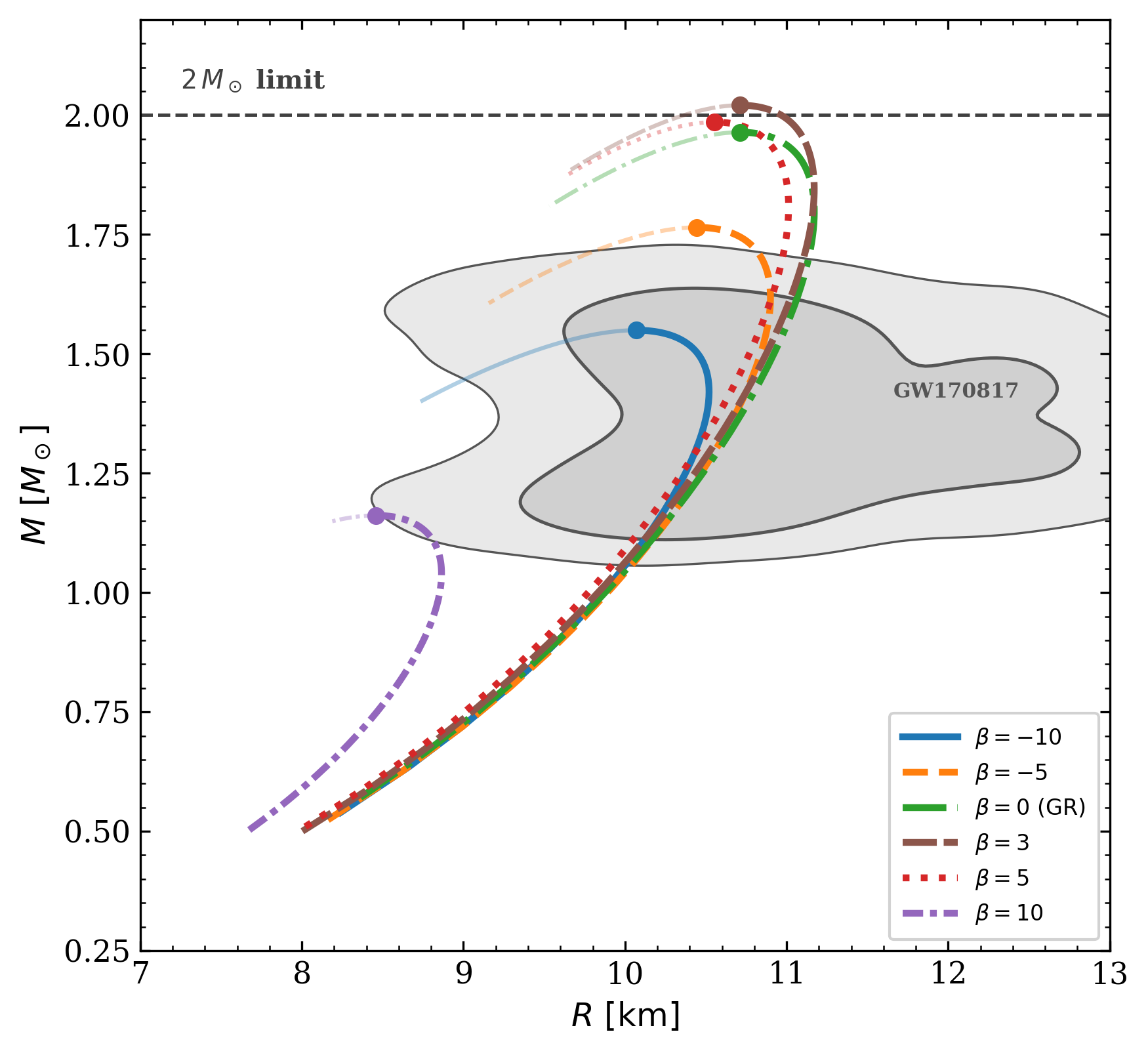}
\caption{Mass--radius relation for quark stars in
$f(T,\mathcal{T})=T+\beta\mathcal{T}$ gravity with the conformal
MIT bag EOS ($B=60\,\mathrm{MeV\,fm^{-3}}$, $\omega=1/3$) for
$\beta\in\{-10,-5,0,3,5,10\}$.
The set includes $\beta=3$, the location of the $\Mmax$ peak
identified in Fig.~\ref{fig:Mmax_beta}, so that the peak
configuration is visible directly in the mass--radius plane.
Solid lines show the candidate stable branch identified by the
turning-point criterion ($dM/d\rho_c>0$); faded lines
show the unstable branch.
Filled circles mark the maximum-mass configuration.
The dashed horizontal line marks the $2\,\Msun$ observational
threshold.
The shaded contours show the GW170817 posterior~\cite{Abbott:2018gw170817}
derived under hadronic-EOS priors; the overlap with quark-star
sequences is shown for orientation only.}
\label{fig:MR}
\end{figure}

\begin{figure}[htbp]
\centering
\includegraphics[width=\columnwidth]{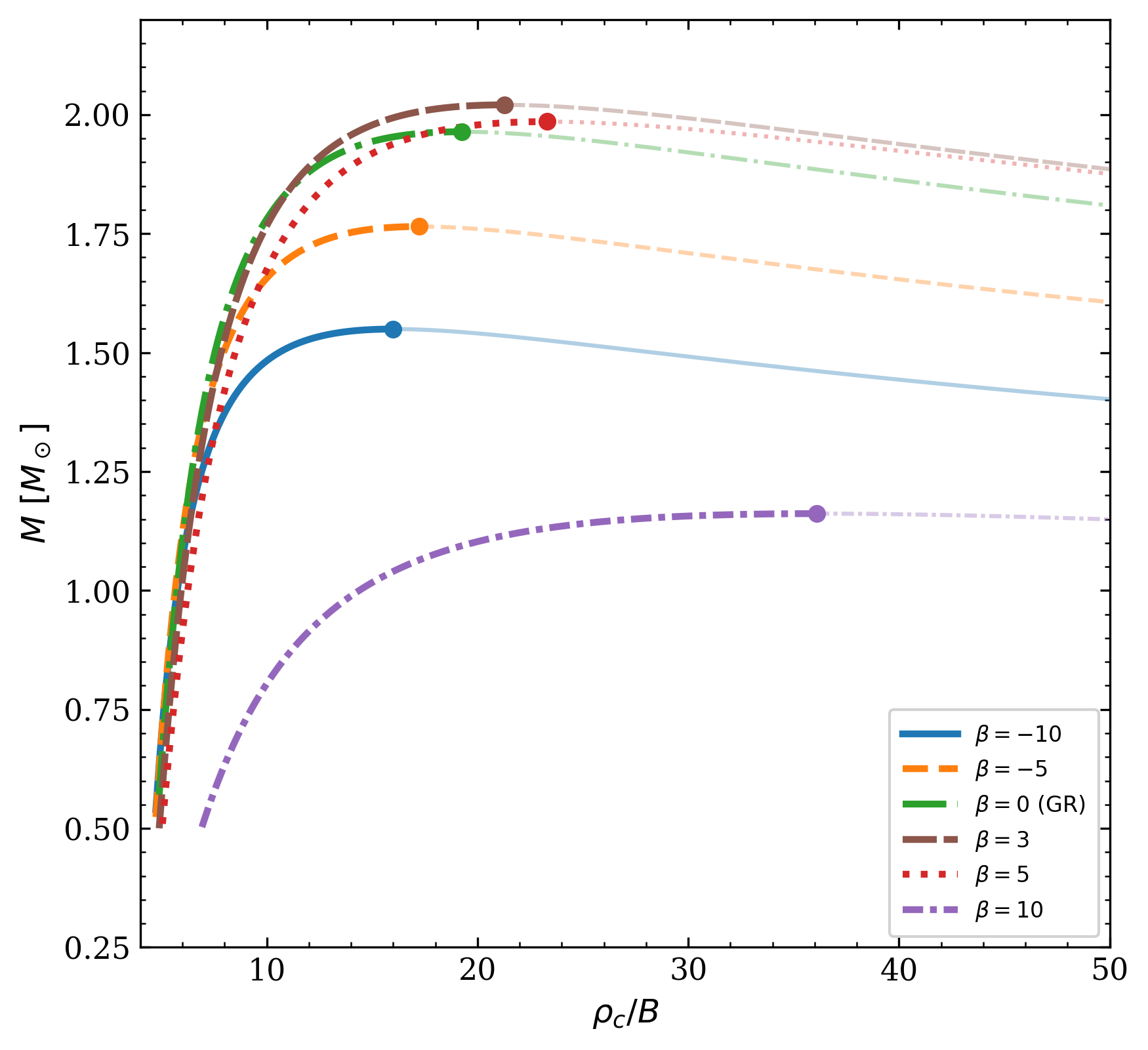}
\caption{Gravitational mass $M$ vs dimensionless central density
$\rho_c/B$ for the same parameter set as Fig.~\ref{fig:MR}.
Solid lines show the candidate stable branch identified by
the turning-point criterion; faded lines show the unstable
branch beyond the turning point (filled circle).}
\label{fig:Mrho}
\end{figure}

The mass--radius sequences for $\beta\in\{-10,-5,0,3,5,10\}$
are displayed in Fig.~\ref{fig:MR}; the corresponding
$M$--$\rho_c/B$ sequences are shown in Fig.~\ref{fig:Mrho}.
Filled circles mark the maximum-mass (turning-point) configuration
on each candidate stable branch; faded lines show the unstable branch
($dM/d\rho_c<0$).

The maximum mass varies non-monotonically with $\beta$; the
asymmetry between the negative and positive branches is examined
in Sec.~\ref{sec:nonmono} and Sec.~\ref{sec:discussion}.
For $\beta<0$ the coupling suppresses $\Mmax$ below the GR value:
$\Mmax=1.549\,\Msun$ at $\beta=-10$ and $1.765\,\Msun$ at
$\beta=-5$ (Table~\ref{tab:stellar_params}).
The GR limit ($\beta=0$) gives $\Mmax=1.964\,\Msun$ at
$R=10.71\,\mathrm{km}$, in agreement with standard MIT bag
results~\cite{Alcock:1986strange}.
At $\beta=3$ the maximum mass rises to $2.021\,\Msun$, exceeding
the $2\,\Msun$ threshold~\cite{Fonseca:2021refined}.
At $\beta=5$ it decreases to $1.985\,\Msun$, and at $\beta=10$
it falls sharply to $1.162\,\Msun$, reflecting the approach to
the singular surface at $\beta=4\pi$.

The central density at $\Mmax$ increases with $\beta$:
$\rho_c^{\max}/B=16.03$ at $\beta=-10$, $19.25$ at $\beta=0$,
$21.25$ at $\beta=3$, and $36.19$ at $\beta=10$
(Table~\ref{tab:stellar_params}).

\subsection{Compactness and surface redshift}

\begin{figure}[htbp]
\centering
\includegraphics[width=\columnwidth]{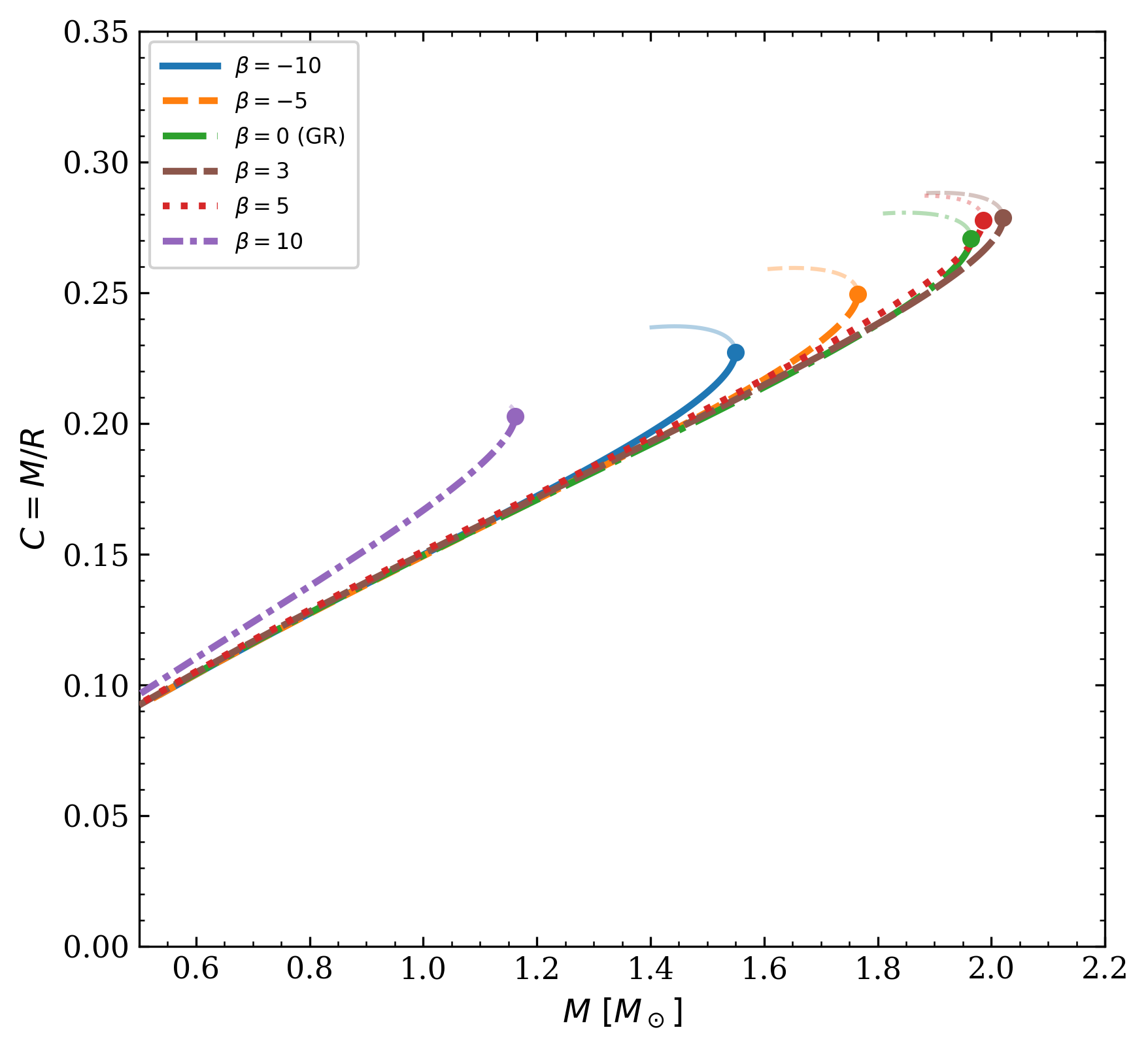}
\caption{Compactness $\mathcal{C}=M/R$ (geometrized units) vs
gravitational mass $M$ along the candidate stable branch
identified by the turning-point criterion for each $\beta$.
Filled circles mark the maximum-mass configuration.
All configurations remain below the standard GR Buchdahl
compactness benchmark, $\mathcal{C}=4/9\approx0.444$.}
\label{fig:compactness}
\end{figure}

The compactness $\mathcal{C}=M/R$ (in geometrized units, $G=c=1$)
along the candidate stable branch is shown in Fig.~\ref{fig:compactness}.
All configurations remain below the standard GR Buchdahl
compactness benchmark, $\mathcal{C}=4/9\approx0.444$~\cite{Buchdahl:1959zz};
the peak compactness is $\mathcal{C}=0.279$ at $\beta=3$.

\begin{figure}[htbp]
\centering
\includegraphics[width=\columnwidth]{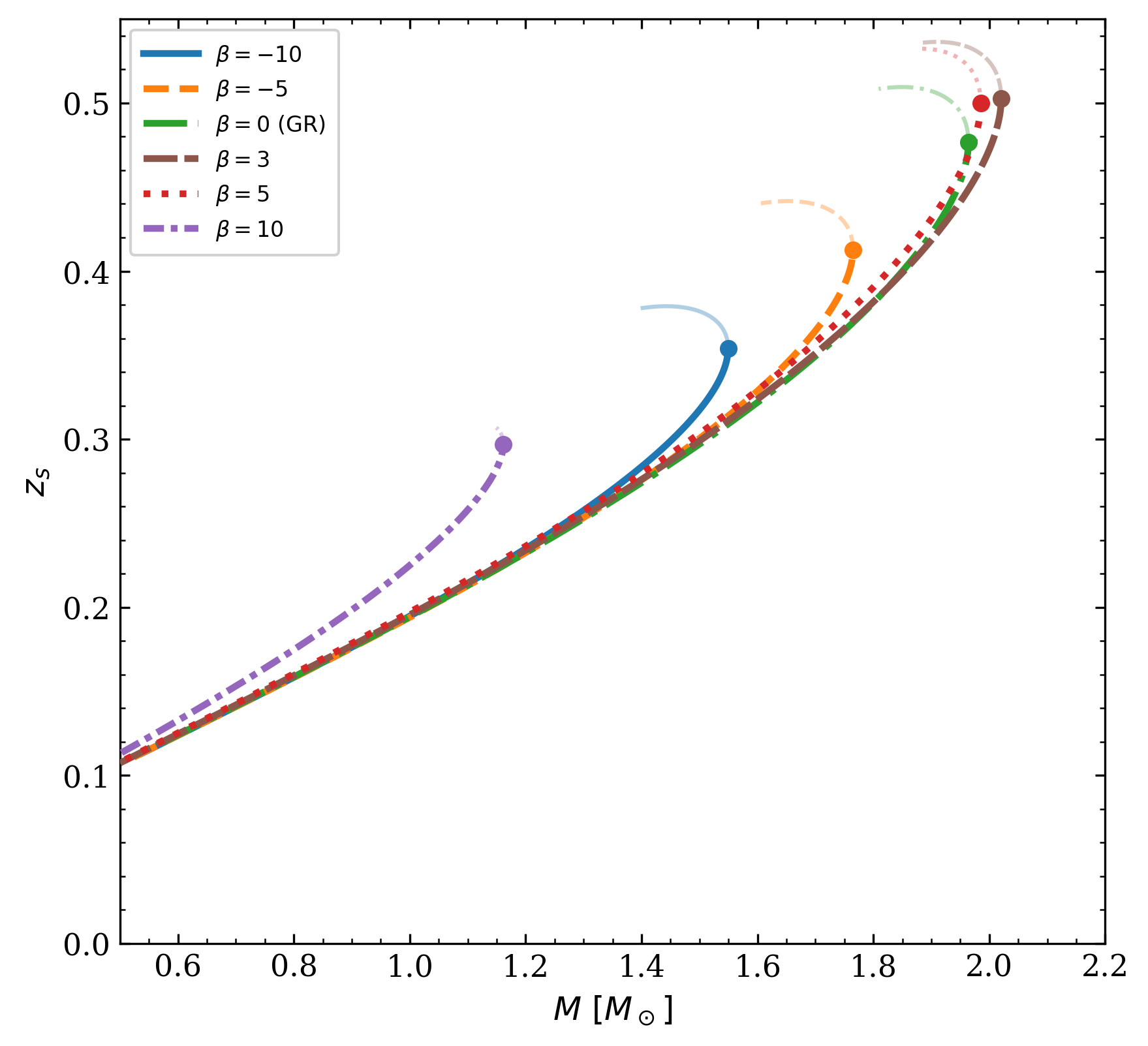}
\caption{Surface gravitational redshift
$z_s=(1-2\mathcal{C})^{-1/2}-1$ vs $M$ along the candidate stable
branch identified by the turning-point criterion.
All configurations remain below the commonly quoted GR reference
value $z_s=0.85$~\cite{Lattimer:2006xb}.
See Table~\ref{tab:stellar_params} for values at $\Mmax$.}
\label{fig:redshift}
\end{figure}

The surface gravitational redshift,
\begin{equation}
z_s=\left(1-\frac{2M}{R}\right)^{-1/2}-1
   =\left(1-2\mathcal{C}\right)^{-1/2}-1,
\end{equation}
is shown in Fig.~\ref{fig:redshift}.
The calculated surface redshifts also remain below the commonly
quoted GR reference value $z_s=0.85$~\cite{Lattimer:2006xb}.
At $\Mmax$ the redshift ranges from $z_s=0.354$ ($\beta=-10$)
to $z_s=0.503$ ($\beta=3$); the full values are in
Table~\ref{tab:stellar_params}.
The enhancement $\Delta z_s=0.026$ at $\beta=3$ relative to GR
is in principle observable through spectroscopic measurements of
surface absorption lines~\cite{Cottam:2002cu}.

\subsection{Non-monotonic dependence of $\Mmax$ on $\beta$}
\label{sec:nonmono}

\begin{figure}[htbp]
\centering
\includegraphics[width=\columnwidth]{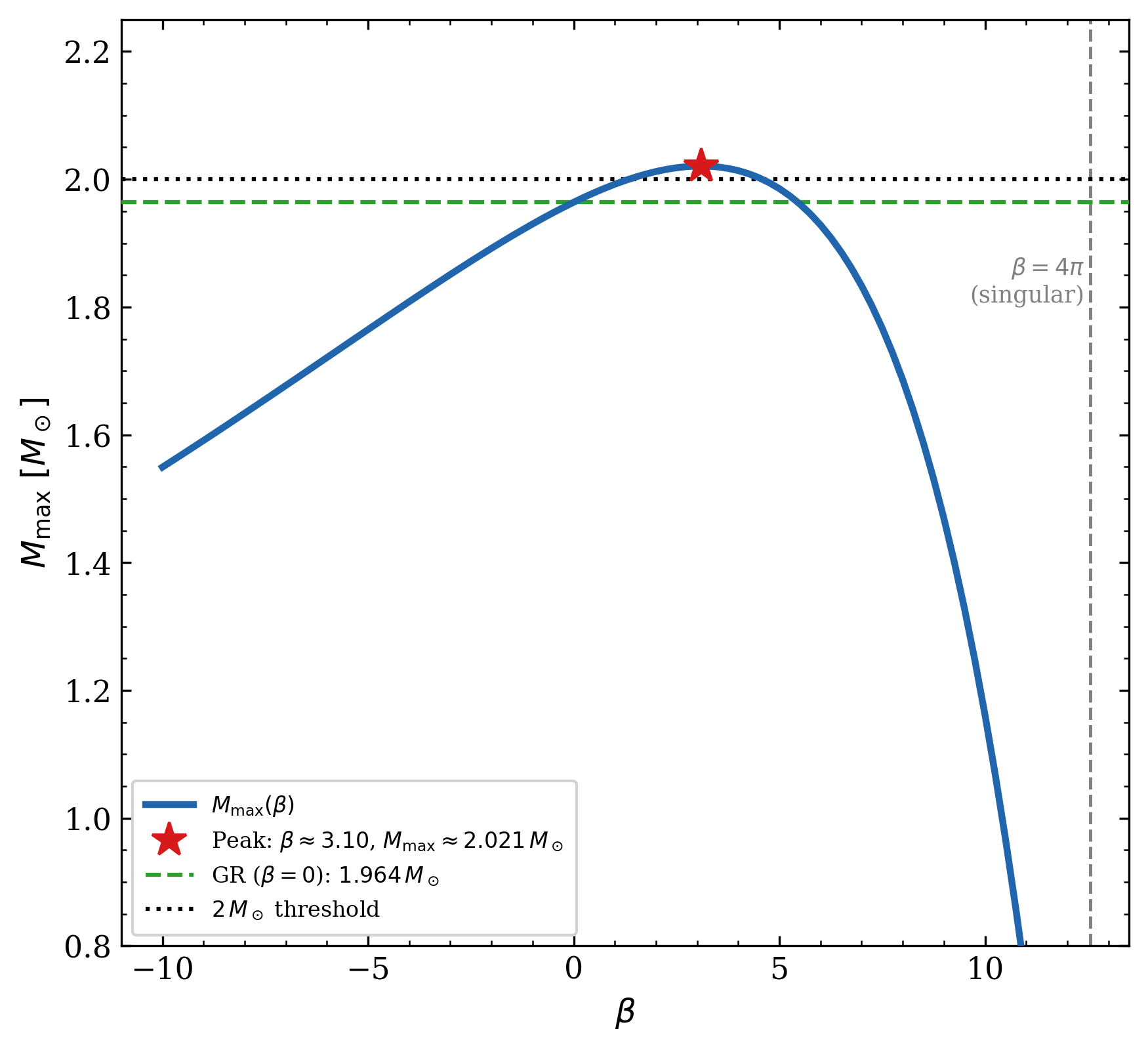}
\caption{Maximum gravitational mass $\Mmax$ vs the trace-coupling
parameter $\beta$ for quark stars in
$f(T,\mathcal{T})=T+\beta\mathcal{T}$ gravity
($B=60\,\mathrm{MeV\,fm^{-3}}$, $\omega=1/3$).
Each point is from a refined turning-point search.
The green dashed line marks the GR reference
($\Mmax^{\rm GR}=1.964\,\Msun$); the black dotted line marks
the $2\,\Msun$ threshold.
The grey dashed vertical line marks the singular surface
$\beta=4\pi\approx12.57$.
The red star marks the peak at $\beta\approx3.10$,
$\Mmax=2.021\,\Msun$, the central result of this work.}
\label{fig:Mmax_beta}
\end{figure}

Figure~\ref{fig:Mmax_beta} shows the result of a dense scan
over $\beta\in[-10,12.25]$ with step $\Delta\beta=0.25$.
The curve rises from $\Mmax=1.549\,\Msun$ at $\beta=-10$,
crosses the GR reference near $\beta=0$, reaches a peak of
$\Mmax=2.021\,\Msun$ at $\beta\approx3.10$, and then decreases
steeply, falling below $2\,\Msun$ near $\beta\approx5.5$ and
plunging toward zero as $\beta\to4\pi^{-}$.

The non-monotonic shape cannot be attributed to a sign change at
$\beta=16\pi/7$.
As shown in Eq.~\eqref{eq:denomfactor}, for $\omega=1/3$ the
denominator of Eq.~\eqref{eq:dpdr} reduces to
$\tfrac{4}{3}(4\pi-\beta)$, whose only zero lies at
$\beta=4\pi$, and the pressure-gradient coefficient
\begin{equation}
\mathcal{K}(\beta)=\frac{(\beta-8\pi)\omega}
{\omega(16\pi-7\beta)+\beta}
=\frac{\beta-8\pi}{4(4\pi-\beta)}
\label{eq:Kcoeff}
\end{equation}
reduces to the general-relativistic value $-1/2$ at $\beta=0$
and grows steadily in magnitude for $\beta>0$, reaching
$1.16$ and $2.95$ times the GR value at $\beta=3$ and
$\beta=10$, respectively.
The coupling therefore does not enhance the maximum mass by
slowing the pressure fall-off.
The non-monotonicity instead emerges from the combined influence
of the trace coupling on the pressure gradient
\eqref{eq:dpdr}, the metric potential \eqref{eq:Aprime}, and the
algebraic radial-metric closure \eqref{eq:closure}, whose
competing effects are resolved only by the numerical solution of
the full coupled system.

The way those effects trade off can nevertheless be read directly
from the coefficients. At $\omega=1/3$ the closure
\eqref{eq:closure} carries
$\Sigma=-2+\beta/6\pi$ and $\xi=\tfrac{10}{3}\beta B_{\rm km}$.
A positive $\beta$ therefore does two things at once: it reduces
$|\Sigma|$ below its general-relativistic value of $2$, weakening the
mass term in $e^{-B_{m}}$, and it adds a positive $\xi$ contribution
proportional to the bag constant. Both push $e^{-B_{m}}$ upward, the
interior metric becomes shallower than in general relativity, and a
given central density supports more mass before the turning point is
reached. Working against this is $\mathcal{K}(\beta)$, which steepens
the pressure fall-off and shrinks the star. For moderate positive
$\beta$ the two closure terms win, which is why $\Mmax$ rises to
$2.021\,\Msun$ near $\beta\approx3.10$ while the radius stays close
to its general-relativistic value (Table~\ref{tab:stellar_params}).
Beyond that point the balance tips. The denominator $4\pi-\beta$ in
\eqref{eq:dpdr} shrinks toward zero, $|\mathcal{K}|$ grows without
bound, and the pressure drops so rapidly with radius that the surface
is reached before much mass has accumulated. The star contracts,
$\rho_c^{\max}/B$ climbs from $21.25$ at $\beta=3$ to $36.19$ at
$\beta=10$, and $\Mmax$ collapses toward zero as
$\beta\to4\pi^{-}$. The turnover is thus the crossing point of a
closure effect that saturates and a pressure-gradient effect that
diverges.

\begin{table}[htbp]
\centering
\caption{Key stellar parameters at the maximum-mass configuration
for $f(T,\mathcal{T})=T+\beta\mathcal{T}$ quark stars with
$B=60\,\mathrm{MeV\,fm^{-3}}$, $\omega=1/3$.
$\mathcal{C}=M/R$ (geometrized units);
$z_s=(1-2\mathcal{C})^{-1/2}-1$.
All values from a refined turning-point search.}
\label{tab:stellar_params}
\begin{tabular}{cccccc}
\hline\hline
$\beta$ & $\Mmax\,(\Msun)$ & $R\,(\mathrm{km})$ &
$\rho_c^{\max}/B$ & $\mathcal{C}$ & $z_s$ \\
\hline
$-10$    & 1.5494 & 10.0655 & 16.030 & 0.22729 & 0.35406 \\
$-5$     & 1.7648 & 10.4433 & 17.245 & 0.24953 & 0.41288 \\
$0$ (GR) & 1.9641 & 10.7121 & 19.249 & 0.27073 & 0.47677 \\
$3$      & 2.0206 & 10.7115 & 21.253 & 0.27854 & 0.50258 \\
$5$      & 1.9853 & 10.5545 & 23.282 & 0.27775 & 0.49989 \\
$10$     & 1.1615 &  8.4581 & 36.189 & 0.20277 & 0.29700 \\
\hline\hline
\end{tabular}
\end{table}

\section{Discussion and Conclusions}
\label{sec:discussion}

We have studied quark stars in
$f(T,\mathcal{T})=T+\beta\mathcal{T}$ gravity with the conformal
MIT bag EOS at $B=60\,\mathrm{MeV\,fm^{-3}}$.
The maximum mass $\Mmax(\beta)$ is non-monotonic in $\beta$: it
peaks at $2.021\,\Msun$ for $\beta\approx3.10$, exceeding the GR
value by $\Delta M=+0.057\,\Msun$ ($+2.9\%$), and is suppressed
toward the singular surface $\beta=4\pi$.
The conventional $2\,\Msun$ pulsar
requirement~\cite{Fonseca:2021refined} is satisfied for
$\beta\in(0,\sim\!5.5)$, so a moderate positive trace coupling is
compatible with this observational threshold within a finite
interval.
The peak mass $\Mmax=2.021\,\Msun$, however, does not reach the
central mass estimate of PSR~J0952$-$0607,
$M\simeq2.35\,\Msun$~\cite{Romani:2022psr}.
Accommodating such a high mass would require either a stiffer
quark-matter equation of state, additional microphysical effects,
or an extension of the present one-parameter model; we note this
as a direction for future work rather than a result established
here.
All configurations on the candidate stable branch are causal,
satisfy $\Gamma>4/3$ analytically, and remain below the standard
GR compactness and surface-redshift benchmarks.
The non-monotonic behaviour cannot be traced to a sign change of
the factor $(16\pi-7\beta)$: for $\omega=1/3$ the denominator of
Eq.~\eqref{eq:dpdr} reduces to $\tfrac{4}{3}(4\pi-\beta)$, and
the pressure-gradient coefficient \eqref{eq:Kcoeff} grows
monotonically in magnitude with $\beta>0$.
The turnover instead reflects the competition between the
coupling's effect on the pressure gradient, on the metric
potential, and on the algebraic closure, and is established here
numerically rather than analytically.

The correspondence established in
Sec.~\ref{subsec:equivalence} places these numbers in a wider setting.
Under $\beta=-2\lambda$ the peak sits at $\lambda\approx-1.55$, the
observationally allowed window becomes $\lambda\in(-2.75,0)$, and the
singular surface is $\lambda=-2\pi$. Read this way the results are a
statement about the negative-coupling branch of $R+2\lambda\mathcal{T}$
gravity as much as about its torsional counterpart, and that branch has
not previously been mapped with a self-bound equation of state. What
the teleparallel formulation contributes is the setting in which the
positive, and physically favoured, side of the coupling is the natural
one to explore.

Some limitations should be kept in mind.
The full radial pulsation equations in $f(T,\mathcal{T})$ gravity
have not been derived; our stability statements rest on the
turning-point criterion~\cite{Harrison:1965} and on the analytic
bound $\Gamma>4/3$~\cite{Chandrasekhar:1964zz}, and a complete
Sturm--Liouville analysis would put them on a firmer footing.
The tidal Love number $k_2$ and the dimensionless deformability
$\Lambda$, which are directly constrained by
GW170817~\cite{Abbott:2017gw170817,Abbott:2018gw170817}, remain to
be computed in this framework.
As in the original compact-star treatment of Pace and Levi
Said~\cite{Pace:2017quarka}, the terminal value of the integrated
mass function has been identified with the gravitational mass of
the star. An explicit analysis of the surface junction conditions
would provide a useful independent test of this prescription, and
is left for future work.
Finally, more realistic quark-matter descriptions, among them
perturbative-QCD corrections~\cite{Fraga:2014interacting,%
Kurkela:2014constraining,S.Fraga:2016neutron} and
colour-superconducting
phases~\cite{Alford:1999colorflavor,Alford:2003compact,%
Lugones:2002colorflavor}, together with a scan over the bag
constant $B$, would broaden the parameter space and sharpen the
comparison with objects such as HESS~J1731$-$347 and the
GW190814 secondary~\cite{Doroshenko:2022strangely,%
Abbott:2020gw190814}.
The unified interacting equation of state, with its repulsive
perturbative-QCD term, is the most promising of these for reaching the
heaviest pulsars~\cite{Zhang:2021unified,Gammon:2024quark}, and a
density-dependent bag parameter offers a further
route~\cite{Saha:2026density}. Whether the coupling window quoted here
survives once the microphysics is stiffened is a question we regard as
open.

In summary, quark stars in $f(T,\mathcal{T})=T+\beta\mathcal{T}$
gravity with the conformal MIT bag EOS display a non-monotonic
maximum-mass profile bounded above by the singular surface at
$\beta=4\pi$, with a peak of $\Mmax=2.021\,\Msun$ at
$\beta\approx3.10$ and observational viability for
$\beta\in(0,\sim\!5.5)$.
These results establish $f(T,\mathcal{T})$ gravity as a viable
framework for massive quark stars, equivalently the negative-$\lambda$
branch of $R+2\lambda\mathcal{T}$ gravity, and motivate further studies of
tidal deformability, radial stability, and extended EOS families.

\begin{acknowledgments}
The authors thank the Editor-in-Chief, the Handling Editor, and the
anonymous Referee for the careful reading of the manuscript and for the
constructive comments, which improved the presentation of this work.
T. Tangphati acknowledges COST action CA21106 and CA22113.
\.{I}. Sakall{\i} is grateful to EMU, T\"UB\.ITAK, ANKOS and
SCOAP3 for their support, and acknowledges the networking support of
COST Actions CA18108, CA22113, CA21106, CA23130, CA23115 and CA21136.

\end{acknowledgments}

\bibliographystyle{apsrev4-2}
\bibliography{finalref}

\end{document}